\documentstyle[preprint,aps,prl,epsf]{revtex} 
\begin{document}
\draft
\tightenlines
\onecolumn
\title{Spin Electronics and Spin Computation}
\author{S. Das Sarma$^1$, Jaroslav Fabian$^{1,2}$, Xuedong Hu$^1$, Igor
\v{Z}uti\'{c}$^1$}  
\address{1. Department of Physics, University of
Maryland, College Park, MD 20742-4111}
\address{2. Max-Planck Institute, Physics of Complex
Systems, Noethnitzer Str. 38, Dresden 01187, Germany}
\date{\today}
\maketitle
\begin{abstract}
We review several proposed spintronic devices that can
provide new functionality or improve available functions of 
electronic devices.  In particular, we discuss a high mobility field
effect spin transistor, an all-metal spin transistor, and our recent
proposal of an all-semiconductor spin transistor and a spin battery. 
We also address some key issues in spin-polarized transport, which are
relevant to the feasibility and operation of hybrid semiconductor
devices. Finally, we discuss a more radical aspect of
spintronic research---the spin-based quantum computation 
and quantum information processing.  
\end{abstract}

\section{Introduction}
\label{section1}

Spintronics, or spin electronics, refers to the study of the role
played by electron (and more generally nuclear) spin in solid state
physics, and possible devices that specifically exploit spin
properties instead of or in addition to charge degrees of freedom
\cite{prinz98}.
For example, spin relaxation and spin transport in metals and
semiconductors are of fundamental research interest not
only for being basic solid state physics issues, but also for the
already demonstrated potential these phenomena have in
electronic technology \cite{prinz98,Wolf,Kikkawa,DFHZ}.  
The prototype device that is already in use in
industry as a read head and a memory-storage cell is the
giant-magnetoresistive (GMR) sandwich structure \cite{prinz98} which
consists of alternating ferromagnetic and nonmagnetic metal layers.
Depending on the relative orientation of the magnetizations in the
magnetic layers, the device resistance changes from small (parallel
magnetizations) to large (antiparallel magnetizations). 
This change in resistance (also called magnetoresistance) is used to 
sense changes in magnetic fields.  Recent efforts in GMR technology
have also involved magnetic tunnel junction devices where the
tunneling current depends on spin orientations of the electrodes. 

Current efforts in designing and
manufacturing spintronic devices involve two different approaches. The
first is perfecting the existing GMR-based technology
by either developing new materials with larger spin polarization of 
electrons or making improvements or variations in the existing devices
that allow for better spin filtering. The second effort, which is
more radical, focuses on finding novel ways of both generation and
utilization of spin-polarized currents. These include investigation
of spin transport in semiconductors and looking for ways in which
semiconductors can function as spin polarizers and spin valves.  The
importance of this effort lies in the fact that the
existing metal-based devices do not amplify signals (although they are
successful switches or valves), whereas semiconductor based
spintronic devices could in principle provide amplification and
serve, in general, as multi-functional devices.  Perhaps even more
importantly, it would be much easier for semiconductor-based devices
to be integrated with traditional semiconductor technology.  

While there are clear advantages for introducing semiconductors
in novel spintronic applications, many basic questions pertaining
to combining semiconductors with other materials to produce
a viable spintronic technology remain open. For example, whether
placing a semiconductor in contact with  another material would
impede spin transport across the interface is far from
well-understood.  In the past, one of the strategies to advance
understanding of spin transport in hybrid semiconductor structures
was to directly borrow knowledge obtained from studies of more
traditional magnetic materials. However, there is also an alternative
approach involving the direct investigation of spin transport
in all-semiconductor device geometries.  In such a scenario a
combination of optical manipulation (for example, shining circularly
polarized light to create net spin polarization) and material
inhomogeneities (e.g. by suitable doping as in the recently
discovered Ga$_{1-x}$Mn$_x$As type ferromagnetic materials where Mn
impurities act as dopants) could be employed to tailor spin transport
properties.

In addition to the near-term studies of various spin transistors and
spin transport properties of semiconductors, a long-term and
ambitious subfield of spintronics is the application of
electron and nuclear spins to quantum information processing and
quantum computation.  It has long been pointed out that quantum
mechanics may provide great advantages over classical physics in
physical computation \cite{Deutsch,Feynman}.  However, the real boom
started after the advent of Shor's factorization algorithm
\cite{Shor1} and quantum error correction schemes
\cite{Shor2,Steane}.  Among the many quantum computer hardwares
that were proposed are the ones based on electron and nuclear spins
\cite{DiVincenzo}.  Obviously, the spins of electrons and spin-1/2
nuclei provide perfect candidates for quantum bits (qubits) as their
Hilbert spaces are generally well-defined and their decoherence
relatively slow \cite{Awschalom}.  

In this paper we review several of the important issues in spintronics
that are mentioned above.  In particular,
in Section \ref{section2} we review some past and
recent attempts at achieving the goals of building practical
spintronic devices.  We discuss perhaps the
first scheme of a spin MOS device, the Datta and Das spin transistor
\cite{datta90} in which current is modulated by spin precession in
the Rashba field controlled by a Schottky barrier voltage. Next, an
all-metal transistor of Johnson \cite{johnson93} is reviewed. In the
Johnson transistor, spin up and spin down states play a similar role
as that of electrons and holes in semiconductor transistors, and the
direction of the current in the working circuit can be switched by
flipping the direction of an applied magnetic field.  Finally, we
describe our recent proposal \cite{zutic01} in designing new schemes
of all-semiconductor spintronic devices, namely, spin-polarized solar
battery, which generates spin-polarized currents without the need of
ferromagnetic electrodes, and magnetic field effect transistor where
an external magnetic (instead of the traditional electric) field
controls the current output.      
In Section \ref{section3} we review some aspects of spin transport
(and more generally spin-polarized transport) in hybrid structures.
This is in part motivated by devices that we discuss
in Section \ref{section2}, some of which include semiconductors.  We
discuss some of our results for spin transport in
semiconductor/superconductor hybrid structures and suggest
possible directions for future research.  
In Section \ref{section4} we briefly review several of the spin-based
quantum computer schemes.  In particular, we discuss the electron spin
based proposals in quantum dots and donors controlled by external
magnetic fields, gates, and by electron-electron exchange or
electron-photon interaction.  We then discuss the nuclear spin based
Si quantum computer proposal and its possible extensions.  Finally, we
discuss a possible source of error in the exchange-based quantum
computer schemes that we recently studied, which demonstrates the
multitude of difficulties one would face in trying to implement any
solid state  spintronic schemes for quantum computation.

\section{Spintronic Devices}
\label{section2}

\begin{figure}
\centerline{
\epsfxsize=4.0in
\epsfbox{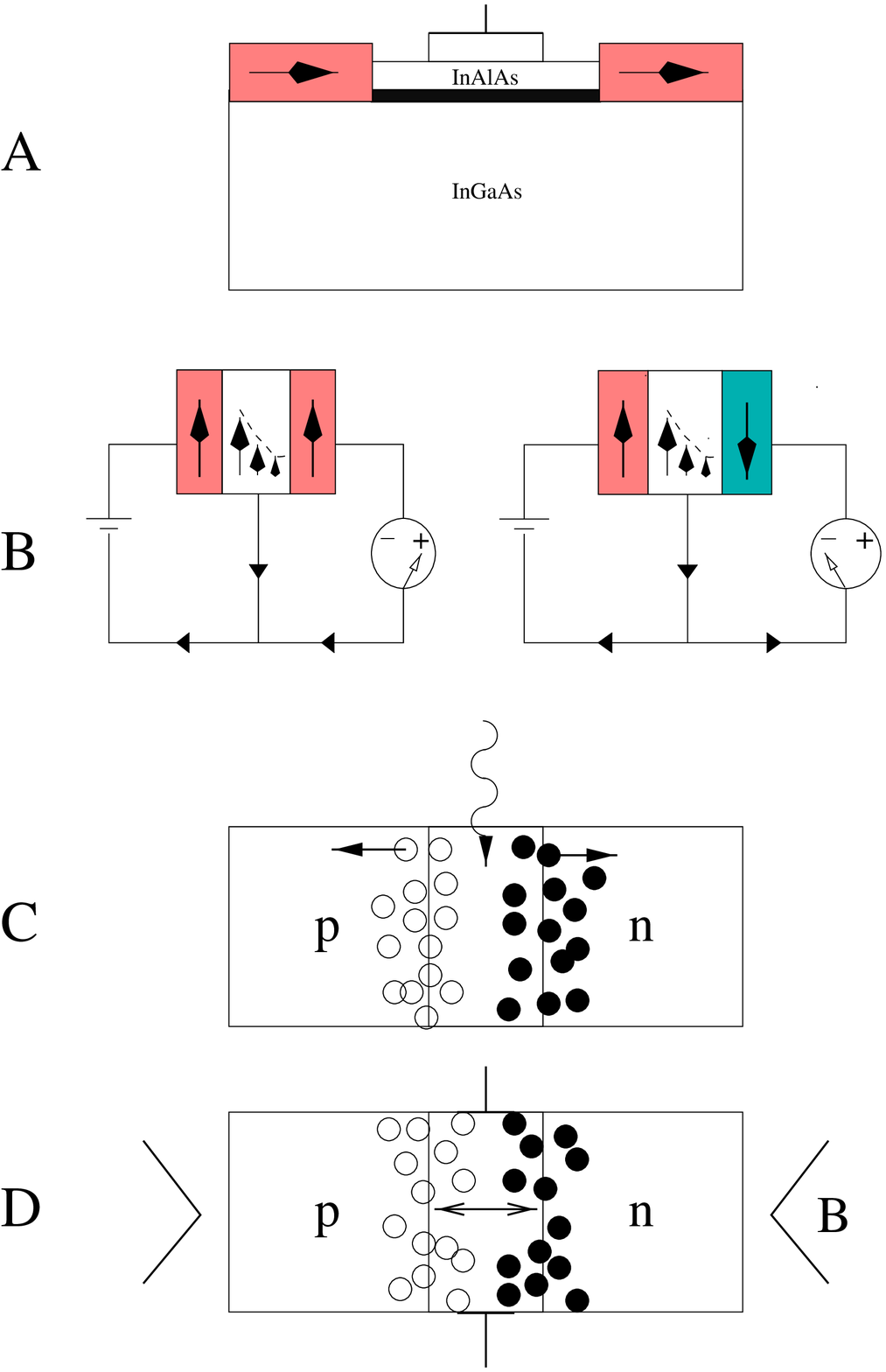}}
\protect\caption[Schemes of selected spintronic devices]
{Schemes of selected spintronic devices. (A) The Datta-Das
spin transistor. Electrons travel in the two-dimensional inverted
region channel (filled region) between two ferromagnetic electrodes.
Electron spins precess in the Rashba field which can be controlled
by the gate voltage, modulating the current.
(B) The Johnson spin transistor.  Depending on the orientation of the
magnetizations in the two ferromagnetic layers, the current in the
collector circuit flows either from the base into the emitter (left)
or from the emitter into the base (right). (C) Spin-polarized solar 
battery. Filtered solar light (circularly polarized) generates 
electron-hole pairs in the depletion region. The polarization is carried 
only by electrons if the semiconductor is III-V, like GaAs. The
resulting current flowing in an external circuit that connects the
$n$ and $p$  regions is spin-polarized. (D) Magnetic field effect
transistor (MFET).  Magnetic field $B$ is applied along the {\it p-n}
junction. The current in the circuit connecting the junction in the
transverse direction depends critically on the size of the depletion
layer (it is small for a larger layer and large for a smaller layer).
If the $g$-factors of the electrons or holes are large, a change in
$B$ can lead to a large change in the width of the depletion layer
and in the magnitude of the transverse current.
}
\label{fig:schemes}
\end{figure}

The first scheme for a spintronic device based on a MOS-like geometry
is the Datta and Das high mobility field effect spin transistor
\cite{datta90}, shown in Fig. \ref{fig:schemes}(A). The
heterostructure (here InAlAs/InGaAs) provides an inversion layer
channel for two-dimensional electron transport between two
ferromagnetic electrodes. One acts as an emitter, the other a
collector.  The emitter emits electrons with their spins oriented
along the direction of  the electrode's magnetization (along the
transport direction in  Fig. \ref{fig:schemes}), while the collector
(with the same electrode magnetization) acts as a spin filter and
accepts electrons with the same spin only.  In the absence of spin
relaxation and spin dependent processes during transport, every
emitted electron enters the collector. The perpendicular field at the
heterostructure interface, however, induces a spin-orbit-like
interaction which acts as an effective (momentum dependent) magnetic
field, in the direction perpendicular to both the transport direction
and the direction of the heterostructure field  (that is, in Fig.
\ref{fig:schemes}(A), perpendicular to the page).  The field
(also called Rashba field) leads to spin precession of the electrons.
Depending on the amount of the electron spin (when entering into the 
collector) 
in the direction of the collector magnetization, the electron current is 
modulated:
an electron passes through if its spin is parallel and does not
if it
is antiparallel to the magnetization. The current is in effect modulated 
by the external electric field induced spin-orbit field naturally
existing in asymmetric zinc blende semiconductor structures (Rashba
effect).  The field can be, in turn, modulated by the applied
perpendicular field at the gate \cite{nitta97}. The Datta-Das
interference  effect should be most visible for narrow-gap
semiconductors like InGaAs which have relatively large spin-orbit
interactions. The effect is yet to be demonstrated experimentally.  

The Johnson spin transistor \cite{johnson93} is a trilayer
structure consisting 
of a nonmagnetic metallic layer sandwiched between two ferromagnets
(for a popular account see Ref. \cite{johnson94}).
It is an all-metal transistor using the same philosophy 
as GMR devices: the current flowing through the structure
is modified by the relative orientation of the magnetic layers
which, in turn, can be controlled by an applied magnetic field. In
this scheme, demonstrated in Fig. \ref{fig:schemes}(B), the battery is
applied in the control circuit (emitter-base), while the direction of
the current in the working circuit (base-collector) is effectively
switched  by changing the magnetization of the collector. The current
is drained  from the base in order to allow for the working current
to flow under the  ``reverse'' base-collector bias (antiparallel
magnetizations). Neither current nor voltage is amplified, but the
device acts as an effective switch or spin valve to sense changes in
an external magnetic field. A potentially significant feature of the
Johnson transistor is that, being all-metallic, it can in principle be
made extremely small using nanolithographic techniques (perhaps
as small as tens of nanometers).  An important disadvantage of
Johnson transistor is that, being all-metallic, it will be difficult
to integrate this spin transistor device into the existing
semiconductor microelectronic circuitry.

An interesting and nontrivial variation of the GMR-like
scheme is a proposal by Monsma et al. \cite{monsma95}
in which the base itself is a metallic nanostructure sandwich of
alternating magnetic-nonmagnetic-magnetic (Co-Cu-Co) layers, while
both the emitter and the collector are semiconductors (Si), providing 
Schottky barrier contacts that allow only hot ballistic carriers to be 
transmitted from the emitter to the collector. As the relative orientation 
of magnetization affects the carrier mean free path (the path 
is smaller for antiparallel orientation), the structure is extremely 
sensitive to external magnetic fields.

A critical disadvantage of metal-based spintronic devices is that they
do not amplify signals. There is no obvious metallic analog of the
traditional semiconductor bipolar transistor (e.g. {\it n-p-n}), in
which draining of one electron from the base allows, say, about
fifty electrons to pass from the emitter into the collector (by
reducing the electrostatic barrier generated by electrons trapped in
the base).  Motivated by the possibility of having both spin
polarization and amplification, we have recently studied a prototype
device, the spin-polarized {\it p-n} junction \cite{zutic01}.  In our
scheme we illuminate the surface of the $p$ region of a GaAs-based
{\it p-n} junction with circularly polarized light to optically
orient minority electrons.  By performing a realistic device modeling
calculation we have discovered that the spin can be effectively
transferred into the $n$ side, via what we call the spin pumping
through the minority channel (in analogy to the optical  spin
pumping in homogeneous semiconductors discovered by D'yakonov
and Perel' \cite{zakharchenya84}).  In effect, the spin gets amplified
going from the $p$ to the $n$ region through the depletion layer.

One application of our proposed spin-polarized {\it p-n} junction is
the  spin-polarized
solar cell \cite{zutic01}, described in Fig. \ref{fig:schemes} (C). 
As in ordinary solar cell batteries, light illuminates the 
depletion layer of a semiconductor (like GaAs), generating electron-hole 
pairs. The
huge built-in electric field in the layer (typically 10$^4$ V/cm) 
swiftly sweeps electrons into the $n$ and holes into the $p$ regions.
If a wire connects the edges of the junction, a current flows.  If
the light is circularly polarized (filtered solar photons), the generated
electrons are spin polarized. (Holes in III-V semiconductors, which
are most useful for opto-spin-electronic purposes, lose their spin
very fast, so that their polarization can be neglected.)  As the 
spin-polarized electrons created in the depletion layer pump the 
spin into the $n$ region, the resulting current is spin polarized.

Finally, Fig. \ref{fig:schemes}(D) shows our recent proposal
of a magnetic field effect transistor (MFET) 
\cite{zutic01}. Electrodes of an external circuit are placed perpendicular 
to the {\it p-n} junction. The current is determined by the 
amount of available electrons in the region of the junction around
the electrodes. If the depletion layer is wider than the electrodes, 
no (or very small) current flows. As the width decreases,
more and more electrons come into contact with the electrodes and the
current rapidly increases. Traditionally, FETs operate with an applied 
electric
field (voltage) along the junction, as the width of the depletion
layer is sensitive to the voltage. We propose to use instead
a magnetic field. If the $n$ or $p$ region
(or both) are doped with magnetic impurities which typically induce
a giant $g$-factor to the current carriers, the magnetic (Zeeman)
energy $g\mu_B B$, where $\mu_B$ is the Bohr magneton, is 
equivalent to having an external voltage of this magnitude. The width of
such a junction could be effectively tailored by an external magnetic
field (differently for spin up and spin down electrons: a spin-polarized
current results as well). Such a device could find use in magnetic
sensor technology like magnetic read heads or magnetic memory cells.

\section{Spin-Polarized Transport}
\label{section3}

The pioneering experiments \cite{tedrow71} on 
spin-polarized transport were performed on 
ferromagnet/superconductor (F/S) bilayers to demonstrate that
current across the F/S interface is spin-polarized. Three decades
later the range of materials where it is possible to study
spin-polarized  transport has significantly increased. Some of the
examples now include  novel ferromagnetic
semiconductors \cite{ohno98}, high temperature 
superconductors \cite{goldman99}, and carbon
nanotubes \cite{balents00}.  Several of the initial questions, such
as the role of interface between different materials and
how to create and measure spin polarization, still remain open and are
of fundamental importance to novel spintronic applications.  We first
turn to the issue of spin transport across interfaces in semiconductor
hybrid devices.  This problem is still not completely resolved and
some of the efforts to understand the remaining puzzles use
analogies with better understood and well-studied charge transport and
current conversion in normal metal/superconductor (N/S) 
structures \cite{andreev64,blonder82,vanson87}. With the effort to
fabricate smaller devices it is possible to reach a ballistic regime
(for example, as in the proposal by Monsma {\it et
al.} \cite{monsma95}) where the carrier mean free path exceeds the
relevant system size and the scattering from interfaces plays a
dominant role. In hybrid structures the presence of magnetically
active interfaces can lead to spin-dependent transmission (spin
filtering) and consequently
influence the operation of spintronic devices by
modifying the degree of spin polarization.

An important case where these ideas are tested is a direct electrical
spin injection from a ferromagnet into a nonmagnetic semiconductor
(Sm). This is also an ingredient needed to implement various proposals
for hybrid semiconductor devices, such as the spin transistor of Datta
and Das \cite{datta90} discussed in the previous section.
In the absence of a complete picture which would describe transport
across F/Sm interface it is helpful to review a simpler unpolarized
case of N/Sm contact.  The charge transport is affected by the
substantial mismatch of carrier densities (or correspondingly Fermi
velocities) and conductivities in the two materials.  Some additional
factors include band bending and pinning of the Fermi level. 
Two generic situations can be distinguished at the N/Sm interface: 
low transparency Schottky barrier and the formation of an accumulation
layer leading to typically higher interfacial transparency. 
These two cases usually correspond respectively to GaAs and InAs
placed in contact with a normal metal.  One would expect an
appropriate spin-dependent generalization of these cases when the
normal metal is replaced by a ferromagnet or if nonmagnetic
semiconductors are replaced by their ferromagnetic counterparts
(Ga,Mn)As and (In,Mn)As.

Reports of spin injection into a semiconductor indicate that the 
obtained spin polarization is substantially smaller than in the 
ferromagnetic spin injector \cite{hammar99}. 
It was suggested \cite{schmidt00},
using the picture of current conversion developed for transport across
the F/N interface in Ref.~\onlinecite{vanson87}, that in the
diffusive regime,  a large mismatch in
conductivities (between the F and the Sm region) presents a basic
obstacle to achieve higher semiconductor spin polarization from
injection.  An interesting proposal was made to circumvent this
limitation.  It was shown that insertion of tunnel contacts between F
and Sm region could eliminate the conductivity mismatch
\cite{rashba00}. To reduce significant material differences between
ferromagnets and  semiconductors alternative methods for spin
injection have concentrated  on using a magnetic
\cite{oestreich99,fiederling99,ohno99} semiconductor as the injector.
While it was shown that this approach could lead to a high degree of
spin polarization \cite{fiederling99} in a nonmagnetic semiconductor,
for successful room temperature spintronic
applications, future efforts will have to concentrate on fabricating
ferromagnetic semiconductors where ferromagnetism will persist at
higher temperatures \cite{medvedkin00}.

\begin{figure}
\centerline{
\epsfxsize=4.0in
\epsfbox{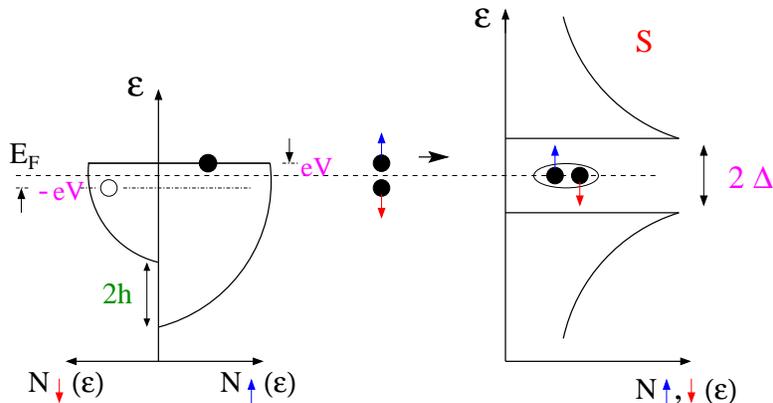}}
\caption{Schematic illustration of Andreev reflection showing the 
density of states in the normal region and in the superconductor
with energy gap $\Delta$,
With spin subband splitting $2h$, only a fraction of incident 
electrons with spin up will be able to find a partner of opposite
spin and contribute to charge transfer by entering superconductor
and  forming a Cooper pair.} 
\label{fig1}
\end{figure}

These issues involving spin injection in semiconductors, 
as well as efforts to fabricate hybrid structures, suggest a need to
develop methods to study fundamental aspects of spin-polarized
transport which are applicable to semiconductors, traditionally
nonmagnetic materials.  In our recent proposal \cite{zutic99} we
suggested studying hybrid Sm/S structures for understanding spin
transmission properties, where the presence of the superconducting
(S) region can serve as a tool to investigate the interfacial
transparency and spin-polarization.  The main motivation is to employ
the two-particle process of Andreev reflection \cite{andreev64}.  In
N/S structures, for a small applied bias V, Andreev reflection is
responsible for current conversion: an incident electron with spin
$\sigma$ slightly above the Fermi energy ($E_F+\epsilon$), together
with an electron slightly below the Fermi energy and of opposite spin
$\overline\sigma$ are transfered into the superconductor where they
form a spin-singlet Cooper pair.  Consequently, the charge of two
electrons is transfered across the interface and normal current is
converted into supercurrent.  The same process can also be viewed as
an incident electron on the N side being reflected at the N/S
interface as a hole accompanied with the transfer of a Cooper pair in 
the S region. It was suggested \cite{beenakker95} that Andreev
reflection would be modified in the presence of spin polarization.
Only a fraction of incident electrons from a majority spin subband
will have partners with opposite spin and consequently can
contribute to charge transfer across the interface through
Andreev reflection \cite{subtle}, as
shown in Fig.~\ref{fig1}.  This was a motivation to develop
experimental methods  based on the conductance measurements to
measure spin polarization \cite{soulen98}.  Theoretical studies have
also considered modifications of spin-polarized transport in the F/S
structures when the S region is a high temperature superconductor
\cite{zhu99,zutic00,kashiwaya99}. Related experiments performed  with
highly polarized ferromagnets suggest that the surface spin
polarization decreases faster with temperature than the
corresponding bulk spin polarization \cite{chen00}.  In using
spin-polarized Andreev reflection to accurately determine the degree
of spin-polarization   some care has to be taken to specify the
appropriate definition of spin polarization \cite{mazin00,geballe00}
and to include the effects of Fermi velocity mismatch
\cite{zutic99,zutic00,geballe00}.

\begin{figure}
\centerline{
\epsfxsize=4.0in
\epsfbox{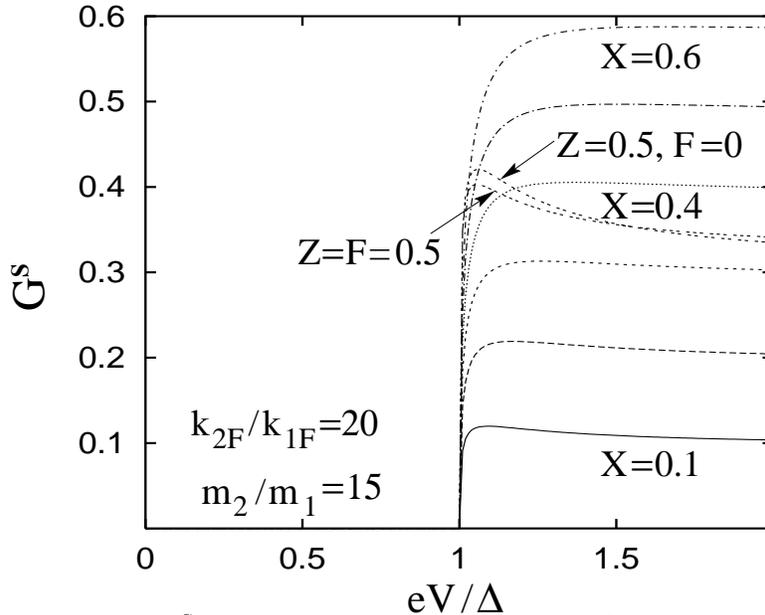}}
\caption[]{Spin conductance $G^{\rm S}(eV/\Delta)$ expressed
in units of $(e^2/h)k_{1F}^2 A/4\pi$, where $k_{1F}$ is the Fermi
wavevector in the Sm region (while $k_{2F}$ corresponds to the S
region). Curves from top to bottom correspond to 
$X\equiv h/E_F=0.6,0.5,0.4,0.3,0.2,0.1$ with no intrinsic interfacial
barrier, except for  two curves representing  $X=0.4$ which 
are labeled by the appropriate interfacial scattering strengths 
$Z$ and $F$ (as defined in Ref.~\protect{\onlinecite{zutic99}})
for potential and spin flip scattering, respectively. 
All the results correspond to the ratio of Fermi wavevectors 
and effective masses (for the S and the Sm region) as denoted in the
figure.} 
\label{fig2}
\end{figure}

In addition to charge transport, which can be used to infer
the degree of spin-polarization, one could also consider
pure spin transport. 
We illustrate this in a hybrid Sm/S structure by 
employing the model and notations from Ref.~\onlinecite{zutic99}. 
We choose a geometry
where semi-infinite Sm and S regions are separated by a flat interface
at which particles can experience potential and spin flip scattering.
In this approach we need to identify the appropriate
scattering processes and the corresponding amplitudes
\cite{blonder82}.  Since all the scattering probabilities should add
up to unity, we can express various quantities of interest only in
terms of scattering processes pertaining to the Sm region.
For example, an electron with spin $\sigma$, incident at the Sm/S
interface  can undergo Andreev reflection (with amplitude $a_\sigma$),
ordinary (potential) reflection (with amplitude $b_\sigma$), 
as well as experience
the corresponding spin-flip process with amplitudes $a_{f\sigma}$
and $b_{f\sigma}$.  Due to the translational invariance along the
interface, the parallel component of the wavevector ${\bf
k}_{\|\sigma}$ is conserved in each scattering process, and by
generalizing the Landauer-B\"{u}ttiker formalism the spin current can
be expressed as \cite{zutic99}
\begin{eqnarray}
I^{{\rm S}}(V,T)=\frac{e}{h} \sum_{{\bf k}_{\| \sigma},\; \sigma}
\int^\infty_{-\infty}
G^{\rm S}_\sigma(\epsilon,k_{{\bf k}_{\| \sigma}})
[f(\epsilon-e V)-f(\epsilon)] d \epsilon, 
\label{gt}
\end{eqnarray}
where $f(\epsilon)$ is the Fermi function,
$G^{\rm S}_\sigma(\epsilon,k_{{\bf k}_{\| \sigma}})
=[1-(v'_{1\overline\sigma}/v_{1\sigma})(|a_\sigma|^2+|b_{f\sigma}|^2)
-|a_{f\sigma}|^2-|b_\sigma|^2] \rho_\sigma$ is the dimensionless 
spin conductance, $\rho_\sigma=\pm 1$ for $\sigma=\uparrow,\downarrow$,
$v_{1\sigma}$ and $v'_{1\overline\sigma}$ 
are the normal components of Fermi velocity before and after 
reflection \cite{zutic99}.  In Fig.~\ref{fig2} we show our calculated
differential spin conductance $G^{\rm S}=dI^{{\rm S}}/dV$ 
as a function of applied bias for different spin polarization,
represented by $X=h/E_F$, where $2h$ is the spin splitting (see
Fig.~\ref{fig1}).  Eq.~\ref{gt} shows that nonvanishing amplitudes for
Andreev Reflection as well as potential and spin flip scattering will
in general reduce the magnitude of the spin current.  This makes sense
since both potential and spin flip scatterings do not
contribute to net transport (charge or spin) across the interface and
Copper pairs transfered into superconductor via Andreev reflection
are spinless \cite{triplet}.   Consequently, spin current is carried
by quasiparticles, and at $T=0$ (as can be seen from a sketch of
density of states in Fig.~\ref{fig1}) the spin current vanishes for
bias less than the superconducting gap. While spin conductance in
Fig.~\ref{fig2} shows high sensitivity to spin polarization, there
remains an experimental  challenge to directly measure the spin,
rather than the usual charge current.

With the recent materials advances of creating spin polarization
(e.g. GaMnAs, InMnAs, etc.) in semiconductors \cite{zutic99} it is
possible to consider various semiconductor based hybrid structures
which in the past have relied on ferromagnets for providing spin
polarization.  For example, there are studies which consider 
heating effects on the transport properties of mesoscopic 
F/S structures \cite{belzig00}, 
or the possibility of implementing switches and logic circuits using 
transitions between normal and superconducting states controlled
by the direction of magnetization in the ferromagnetic
region \cite{kulic00}.  One of the implementations could be to
consider replacing a conventional ferromagnet by a Mn-doped
ferromagnetic semiconductor.

\section{Spin-based Quantum Computation}
\label{section4}

One of the most ambitious spintronic devices is the spin-based
quantum computer (QC) in solid state structures (see \cite{review}
and references therein).  Using electron (or nuclear) spin for
QC purposes is a manifestly obvious idea since a fermion with spin
1/2 is a natural and intrinsic qubit.  Quantum computation requires
both long quantum coherence time and precise external control
\cite{DPD3}. Because of the requirement of very long coherence time
for a QC, both nuclear spin and electron spin have been proposed as
qubit in a QC \cite{Awschalom,review1}.  Since more and more schemes
are being proposed (for example, see \cite{Privman,Barnes}), we will
not attempt a complete review of the field.  Instead, we only review
some of the representative schemes proposed during the past several
years, and discuss mostly our own recent work on electron spin based
quantum computation.   

One of the earliest proposed solid state QC schemes uses the spin of
a single electron trapped in a quantum dot as its qubit 
(see \cite{DPD1,LD,DPD2,BLD,HD1} and references therein).  Local
magnetic fields are used to manipulate single spins, while inter-dot
exchange interaction is used to couple neighboring qubits and
introduce two-qubit entanglement.  A single trapped electron in
a quantum dot implies an extremely low carrier density, which means
very low spin-orbit coupling as the electrons occupy states at the
bottom of the GaAs conduction band and have essentially S type states
\cite{HD1}.  Thus the electron spin coherence time should be much
longer than in the bulk.  However, to trap a single electron in a
gated quantum dot is a difficult task experimentally.  In addition,
to apply a local magnetic field on one quantum dot without affecting
other neighboring dots and trapped spins may also be impossible in
practice.  To overcome the potential problem of local field, an
exchange-based QC model has recently been proposed \cite{DPD4}, which
uses solely the exchange interaction between nearest neighbors to
fulfill both single and two-qubit operations. Here qubits are
combinations of single spin quantum dots (or other basic units such
as donor and nuclear spins) which form the so-called
decoherence-free subspace \cite{Lidar,Bacon}.  Regarding the
difficulty of trapping single electrons in an array of quantum dots,
we recently have showed through a multi-electron calculation that an
odd number of trapped electrons in a quantum dot can be effectively
used as a qubit subject to certain conditions \cite{HD2}.  In
addition to the above operational problems of this QC proposal, there
is still the question of how to reliably measure single electron
spins (or two-spin states).  Various proposals have been put
forward \cite{DPD1,LD,DPD2,Kane2,RSL}, while an extensive experimental
exploration is still needed for any consensus to emerge.  No
experimental results on this issue have yet been reported in the
literature.

The major role played by quantum dots in the above proposal is
to provide tags for individual qubits through a parabolic confinement
of the individual electrons,
thus donor nuclei is a natural alternative to quantum dots.  Indeed,
such a scheme has been proposed \cite{Vrijen}, although it was
originally motivated by the nuclear spin based silicon QC proposal
that we will discuss below.  In this scheme the variation of
g-factor due to varying composition in SiGe alloys is used
together with external gates to selectively provide single-qubit
operations.  Two-qubit operations are again provided by the exchange
interaction between neighboring donor electrons and are controlled by
a combination of external gates and variation of the g-factor.

Another variant of the quantum dot QC is a combination of quantum dot
trapped electron spin and semiconductor microcavity \cite{Imam}. 
Here a single cavity photon in the whispering gallery mode plays the
role of intermediary between two quantum dots.  The
self-assembled quantum dots are embedded in a microdisk cavity.
Each of the dots is doped by one extra electron and is addressed
individually by lasers from fiber tips using near-field techniques.
Single-qubit operations can be achieved through a Raman 
coupling of the spin up and down states of the conduction electron
by using two laser beams from the fiber tips.  Two-qubit operations
are based on cavity-photon-mediated Raman transition for the two
relevant spins analogous to the atomic cavity QED schemes
\cite{Pellizzari}.  Since this scheme uses external laser fields
extensively, the relation between a high Q cavity and all the coupling
to external fields has to be dealt with carefully.  On the other hand,
coherent control has been most successfully demonstrated with light,
so a photon-mediated QC scheme should certainly be seriously
explored.

One of the most intriguing and influential QC schemes is the 
nuclear spin based Si QC \cite{Kane1}.  Here spin-1/2 donor nuclei
are qubits, while donor electrons together with external gates
provide single-qubit (using external magnetic field) and two-qubit
operations (using hyperfine and electron exchange interactions).
Here donor electrons are essentially shuttles between different
nuclear qubits and are controlled by external gate voltages.  In
addition, the final measurement is also over the donor electrons by
converting spin information into charge information \cite{Kane2}.  A
significant advantage of silicon is that its most abundant isotope is
spinless, thus providing a ``quiet'' environment for donor nuclear
spin qubit.  In general, nuclear spins have very long coherence
times because they do not strongly couple with their environment,
and are thus good candidates for qubits.  However, this isolation from
the environment also brings with it the baggage that individual
nuclear spins are difficult to control.  This is why donor electrons
play a crucial role in the Si QC scheme.  Another potential advantage
of a QC based on Si is the prospect of using the vast resources
available from the Si-based semiconductor chip industry. In addition,
the exchange-only schemes can also be applied here, with hyperfine
and electron exchange interaction together providing the nuclear spin
exchange interaction.

One attempt to overcome problems of bulk solution NMR QC
and to reproduce their successes involves using planes of spins
in a crystal lattice \cite{Fumiko}.  In this scheme an inhomogeneous
magnetic field is used to differentiate atomic planes in a
lattice.  Within each plane the spins form a mini-ensemble, leading
to the possibility of producing sufficiently strong signal for
measurement. Furthermore, in such a scheme nuclear spins can be
initialized, thus  overcoming the worst problem in bulk solution NMR
schemes---the unavoidable ensemble average.  Since Maxwell equations
dictate that magnetic field cannot vary linearly along one direction
while remaining uniform along the other two directions, the
equal-field surface must have certain curvature, which means that
only part of the spins in a plane contribute to the signal. 
Nevertheless, even if this particular scheme may
turn out to have intractable experimental difficulties, the
idea of combining NMR spectroscopy and nanostructure manipulation
is certainly worth pursuing further. 

Aside from exploring various schemes to utilize spins for the purpose
of qubits in quantum computation, an equally important task is to
clarify the type of errors in spin-based QCs and how they can be
corrected.
For example, one possible error in the two-qubit operations of the
exchange-based spin QCs is caused by inhomogeneous magnetic
fields \cite{HDD}.  Such a field may come from magnetic impurities or
unwanted currents away from the structure.  Magnetic field affects
both  orbital and spin part of the electron wavefunction.  The orbital
effect is accounted for by adjusting the exchange coupling J, while
the spin effect is accounted for through Zeeman coupling terms:
\begin{eqnarray}
H_s & = & J({\bf B}) {\bf S}_1 \cdot {\bf S}_2 + \gamma_1 S_{1z} 
+ \gamma_2 S_{2z} \,,
\label{eq:spin-Hamiltonian} 
\end{eqnarray}
where ${\bf S}_1$ and ${\bf S}_2$ refer to the spins of the two
electrons; $J({\bf B})$ is the exchange coupling (singlet-triplet
splitting); $\gamma_1$ and $\gamma_2$ are the effective strengths of
the Zeeman coupling in the two quantum dots.  In an inhomogeneous
field, $\gamma_1 \neq \gamma_2$, so that the Zeeman terms do not
commute with the exchange term in the Hamiltonian 
(\ref{eq:spin-Hamiltonian}).  We have done a detailed analysis on
how to achieve swap with such a Hamiltonian, and found that there is
at the minimum an error proportional to the square of field
inhomogeneity in the swap.  For example, if the initial state of the
two electron spin is $|\phi(0)\rangle=|\!\uparrow \downarrow \rangle$,
the density matrix of the first spin after the optimal swap
is
\begin{equation}
\rho_1=\frac{1}{1+x^2}|\!\downarrow\rangle \langle \downarrow\!|
+ \frac{x^2}{1+x^2}|\!\uparrow\rangle \langle \uparrow\!| \,,
\label{eq:swap_error}
\end{equation}
where $x=(\gamma_1-\gamma_2)/2J$.  In other words, 
the first spin can never exactly acquire the state ($|\!\downarrow 
\rangle$) of the second spin.  Its state will remain mixed and
the smallest error from an exact swap is $x^2/(1+x^2)$,
which needs to be corrected.  We have estimated \cite{HDD} that in
GaAs a Bohr magneton can lead to an error in the order of $10^{-6}$,
which is within the capability of currently available quantum
error correction schemes.

\section{Acknowledgement}

This work is supported by US-ONR, DARPA, ARDA, and LPS at the
University of Maryland.


\begin{thebibliography}{00}
\bibitem{subtle} To better understand various
scattering process, in a ballistic regime, it is helpful to use 
an optical analogy and Snell's law\protect{\cite{zutic99,zutic00}}. 
Consequently, depending on the incident angle and spin subband 
splitting\protect{\cite{zutic99,zutic00,mazin00}},
there are cases with no charge transport across the interface.
\end{thebibliography}
\end{document}